\documentclass{article}

\usepackage{arxiv}
\usepackage{amsmath}

\usepackage[utf8]{inputenc} 
\usepackage[T1]{fontenc}    
\usepackage{url}            
\usepackage{booktabs}       
\usepackage{amsfonts}       
\usepackage{nicefrac}       
\usepackage{microtype}      
\usepackage{lipsum}		
\usepackage{graphicx}
\usepackage{doi}

\newcommand{\figref}[1]{Fig.\ \ref{#1}}

\DeclareMathOperator*{\argmaxA}{arg\,max}

\usepackage{hyperref}       

\title{Some Challenges in Monitoring Epidemics}

\date{} 					

\author{
	{Vaiva Vasiliauskaite$^*$}\\
	Computational Social Science, ETH Zurich, Zurich, Switzerland \\
	\And  
	Nino Antulov-Fantulin$^{*,**}$ \\
	Computational Social Science, ETH Zurich, Zurich, Switzerland \\
	\texttt{anino@ethz.ch} \\
	\And
	Dirk Helbing \\
	Computational Social Science, ETH Zurich, Zurich, Switzerland \\
	Complexity Science Hub Vienna\\[4mm]
	$^*$ Shared first authorship\\
	$^{**}$ Corresponding author
}

\begin{document}
\maketitle

\begin{abstract}
Epidemic models often reflect characteristic features of infectious spreading processes by coupled non-linear differential equations considering different states of health (such as Susceptible, Infected, or Recovered). This compartmental modeling approach, however, delivers an incomplete picture of the dynamics of epidemics, as it neglects stochastic and network effects, and also the role of the measurement process, on which the estimation of epidemiological parameters and incidence values relies. In order to study the related issues, we extend established epidemiological spreading models with a model of the measurement (i.e. testing) process, considering the problems of false positives and false negatives as well as biased sampling. Studying a model-generated ground truth in conjunction with simulated observation processes (virtual measurements) allows one to gain insights into the limitations of purely data-driven methods to assess the epidemic situation. 
We conclude that epidemic monitoring, simulation, and forecasting are wicked problems, as applying a conventional data-driven approach to a complex system with non-linear dynamics, network effects, and uncertainty can be misleading. Nevertheless, some of the errors can be corrected for, using scientific knowledge of the spreading dynamics and the measurement process. We conclude that such corrections should generally be part of epidemic monitoring, modeling, and forecasting efforts. 
\end{abstract}


\section{Introduction}

Since thousands of years, the epidemic spreading of diseases has been one of the greatest threats to societies around the world. Now, with increased mobility, diseases tend to spread faster and wider due to air traffic and, in fact, often globally. The resulting pandemics can cost the lives of millions of people and disrupt the social, economic, public, political, and cultural life. 
\par
These challenges are countered with new measurement and vaccination technologies \cite{VDD20}. In particular, digital technologies have enabled data-driven and AI-based methods \cite{AWSCCGCDKG20,VJKH20,bharadwaj2021computational}, which have become quite popular. Some organisations envision a future with ubiquitous health measurements, perhaps even using in-body sensors \cite{eltorai2016microchips}. In any case, a data-driven approach is now common. It often assumes that the overall picture of the situation will become more accurate with a greater amount of data. 
\par
According to Chris Anderson's view of Big Data, the data deluge will make the scientific method obsolete \cite{ChrisAndersonWiredMagazine2008}. If we just had enough data, it would reveal the truth by itself. It would show us where the problems are and what we had to do to fix them. This point of view, however, has been questioned.
Conventional Big Data analytics suffers from a variety of possible problems, including the following:
    
    (i) If there are correlations between two variables $A$ and $B$, it is not clear whether $A$ causes $B$ or $B$ causes $A$, or whether a third variable $C$ causes $A$ and $B$, while this matters a lot for the effectiveness of measures taken. For recent advances
    in causal inference see \cite{pearl2009causal}.
    
    (ii) Detected patterns in the data are not always meaningful~\cite{S13}.
    They can be accidental, and the patterns detected results of overfitting \cite{H03,Y19}. Spurious correlations are, in fact, a wide-spread problem \cite{calude2017deluge}. An appropriate understanding of statistical learning theory is, therefore, crucial \cite{hastie2009elements}.
    
    (iii) Measurements may be biased or the investigated sample non-representative. Therefore, patterns learned may be biased, too. This causes issues in many applications of AI~\cite{L18b}. For example, undesirable discrimination of women, people of color, and minorities has often been found, when machine learning is applied to data of job applicants or face recognition \cite{RGMBLD20,L18}. However, biases are also relevant for testing strategies and modeling of epidemics~\cite{bottcher2021TestStatistics,bentley2021error,campbell2020bayesian}. 
    
    (iv) Classification errors are generally an issue. In particular, for almost every measurement method, there are errors of first kind and errors of second kind, i.e. false positives (''false alarms'') and false negatives (''alarms that do not go off'') \cite{colquhoun2014investigation,banerjee2009hypothesis}. As a result, the sensitivity of a measurement may be high, but its precision may be quite limited or even disappointing. A highly sensitive method would avoid that alarms don't go off, but may cause many false alarms. This is, for example, the case for many ''predictive policing'' approaches, where there are thousands of false positives for every real terrorist. A similar problem is expected to occur in connection with the tracing of potentially infected people, as infection is a probabilistic process. For example, when using mass surveillance tools to determine likely infections~\cite{ram2020mass}, the expected outcome is that many healthy people will be put in quarantine \cite{do2020covid}.   
    
    (v) The measurement of model parameters will be always possible with a finite precision only. Consequently, the exact parameters may never be known, but rather a ``confidence interval''. Any of the values in the confidence interval could be the true parameter. However, the implications for these different values may be pretty different, which limits the accuracy particularly in case of sensitive parameter dependencies. 
    
    (vi) In complex, networked systems with probabilistic behavior and cascading effects (such as epidemic spreading), ``fat-tailed distributions'' pose additional challenges to data analytics and machine learning~\cite{taleb2020single,beirlant2006statistics}. 

In the following, we will investigate, to what extend some of the above issues may undermine an accurate assessment of the state of epidemics, even if a large amount of data is available. This is important in particular, as it is increasingly common to respond to measurement-based predictions rather than to current data, in order to take proactive measures. As we will discuss, this can have undesirable side effects. One response to an anticipated increase of infections may be to engage in more testing to capture the expected rise. The motivation for this is clear: if infections are underestimated, hospitals might not be able to handle the number of emergencies, while an overestimation may lead to unnecessary lockdowns with severe socio-economic consequences. In both cases, unnecessary loss of lives may occur. However, is it always better to make more tests?
\par
The precondition for accurate predictions is to have reliable methods to judge the actual epidemic situation well. The currently used epidemic modeling methods try to describe the disease dynamics by interacting ``agents'' on a population, meta-population, or individual level~\cite{vespignani2020modelling, engbert2021sequential, maier2020effective, jia2020population, siegenfeld2020opinion}. Such models are, for example, used to assess the influence of population density, demographic factors, mobility, or social interactions on the actual disease dynamics~\cite{van2011gleamviz}. Data-driven or machine learning models make fewer assumptions about the actual dynamics and are applicable to a broader range of prediction problems, but they come at the cost of less explainability. Of course, it is also possible to combine classical modeling approaches with machine learning \cite{wu2021deepgleam}.
\par
However, it seems that in many policy decisions, issues related to measurement processes are not yet sufficiently considered. Our contribution, therefore, highlights problems related to monitoring epidemics through measurement processes. As it turns out, it is dangerous to assume that data-driven approaches would be largely exact, or that more measurements or tests or data would automatically give a better picture. 
In the following, we will demonstrate the possible pitfalls of such an approach by analysing the estimation errors resulting from measurement errors, mean value approximations, randomness and network interactions. While certain corrections for the effect of false positives and false negatives have been proposed before~\cite{bottcher2021TestStatistics,campbell2020bayesian,bentley2021error,catala2021robust,wu2020substantial}, here we present a framework that adds a measurement model to a model of epidemic dynamics, considering also stochastic and network effects. We further discuss how to correct for biases with mean-field and Bayesian approaches and what are the fundamental limitations in estimating the state of epidemics.
\section{Epidemic Models}

Depending on the characteristics of a disease, there are different compartmental models in epidemiology to reflect the spreading of a disease and the recovery from it by coupled differential equations. These are mean value equations implicitly assuming that the infection process is well characterized by averages and that correlations do not matter. In the following, we will present three common examples of such epidemiological spreading models.

\subsection{SIR Model}

The SIR model \cite{kermack1927contribution,anderson1992infectious} assumes that all people recover from the disease and are immune after recovery. It distinguishes Susceptible, Infected, and Recovered. Intuitively, their numbers at time $t$ are represented by $S(t)$, $I(t)$, and $R(t)$. The increase in the number of Infected is proportional to the number of Susceptible and the number $I$ of those who can infect them. In the notation used below, the proportionality constant is $b$. Infected recover at a rate $c$. Hence,  
the differential equations describing the change of their numbers in the course of time are 
\begin{eqnarray}
    \dot{S} &=& -bSI \, , \\
    \dot{I} &=& bSI - cI \, , \\
    \dot{R} &=& cI \, ,
\end{eqnarray}
where we use the notation $\dot{Z} = dZ/dt$. Moreover, we have the normalization equation
\begin{equation}
    S(t) + I(t) + R(t) = N \, , 
\end{equation}
where $N$ is the number of people in the population.

\subsection{SEIR Model}

The SEIR model~\cite{anderson1992infectious} assumes that all people recover from the disease, but are immune after recovery. Besides Susceptible, Infected, and Recovered, it considers an Exposed category representing people who have caught the disease, but are not infectious, yet. Intuitively, the numbers at time $t$ are represented by $S(t)$, $I(t)$, $R(t)$, and $E(t)$. For simplicity, we will assume that the Exposed do not have symptoms of the disease (yet), i.e. they (erroneously) appear to be healthy. The increase in the number of Exposed is proportional to the number of Susceptible and the number $I$ of those who can infect them. The proportionality constant is $b$. After some time, Exposed turn into Infected (with symptoms). This happens at a rate $a$. Infected finally recover at a rate $c$. Hence, the coupled differential equations describing the change of their numbers in the course of time are:
\begin{eqnarray}
    \dot{S} &=& -bSI \, , \\
    \dot{E} &=& bSI - aE \, , \\
    \dot{I} &=& aE - cI \, , \\
    \dot{R} &=& cI \, . 
\end{eqnarray}
Moreover, we have the normalization equation
\begin{equation}
    S(t) + E(t) + I(t) + R(t) = N \, .  
\end{equation}

\subsection{SEIRD Model}

The SEIRD model~\cite{anderson1992infectious} assumes that some people recover from the disease and are immune after recovery, while some people die. Besides Susceptible, Exposed, Infected, and Recovered, it considers a Died category. Intuitively, the numbers at time $t$ are represented by $S(t)$, $E(t)$, $I(t)$, $R(t)$ and $D(t)$. The increase in the number of Exposed is proportional to the number of Susceptible and the number $I$ of those who can infect them. The proportionality constant is $b$. Exposed turn into Infected with a rate $a$. Infected recover with a rate $c$ and die with a rate $d$. The differential equations describing the change of their numbers in the course of time are:
\begin{eqnarray}
    \dot{S} &=& -bSI \, , \\
    \dot{E} &=& bSI - aE  \, , \\
    \dot{I} &=& aE - cI - d I\, , \\
    \dot{R} &=& cI \, , \\
    \dot{D} &=& dI  \, . 
\end{eqnarray}
Moreover, we have the normalization equation
\begin{equation}
    S(t) + E(t) + I(t) + R(t) + D(t) = N \, .  
\end{equation}

\subsection{Measuring the State of an Epidemic}

The underlying question of this paper is: how well can the state of epidemics be measured, using common test and monitoring methods? When discussing this, we must consider that all test methods have false positive rates and false negative rates \cite{banerjee2009hypothesis}. This also applies to virtual measurement methods such as those based on inference from tracing data. We assume that Infected are being tested with probability $p$ and people without symptoms ($S$, $E$, and $R$) with probability $q$. When the testing is focused on infected people with symptoms, we will have $q\le p$ or even $q\ll p$. 
\par
In the following, for the sake of illustration, we will focus on the SEIRD model as ground truth. Furthermore, let the false positive rate (FPR) when healthy people are tested be $\alpha$, and the false negative rate (FNR) be $\beta$ when Infected are tested, but $\beta'$ when Exposed are tested. Under these circumstances, the expected number of positive tests is
\begin{equation}
  N_p(t) = (1-\beta) p I(t) + (1-\beta') q E(t) + \alpha q [S(t)+R(t)] \, . 
  \label{ex}
\end{equation}
From this formula, we can immediately see that $N_p(t)$ can be assumed to be proportional to the number $I(t)$ of Infected only if $\beta'=1$ and $\alpha=0$, which is unrealistic, or if $q=0$, i.e. ``healthy'' people are not tested. Otherwise, $N_p(t)$ can be quite misleading. In fact, the term $\alpha q (S+R)$ may dominate the other terms, if a large number of people without symptoms is being tested. In such a case, most positive tests would be false positives, and the number of positive tests would increase with the overall number $N_T$ of tests made. Testing a large number of people without symptoms might, therefore, produce pretty arbitrary and misleading numbers. As a consequence, the number of positive tests is not a good basis for policy measures. In particular, if the testing rate $q$ is increased with the anticipated number of positive tests $N_p(t+\Delta t)$ at some future point in time $t+\Delta t$, this may lead to a ``lock-in effect''. Then, for a long time, politics may not find a way out of the vicious cycle of more predicted cases triggering more tests, which implies more measured and predicted cases, and so on. 
\par
It is much better to compare the number $N_p$ of positive tests given by Eq. \eqref{ex} with the overall number of tests
\begin{equation}
   N_T(t) = p I(t) + q [S(t)+E(t)+R(t)] \, .
\end{equation}
Accordingly, the estimated (measured) proportion of infected people is 
\begin{equation}
\fbox{ $ \displaystyle 
   \hat{s}(t) = \frac{N_p(t)}{N_T(t)} = \frac{(1-\beta) p I + (1-\beta') q E + \alpha q (S+R)}{p I + q (S+E+R)} $}  
   \label{success}
\end{equation}
Here, we have dropped the argument $(t)$ on the rhs for simplicity. Let us now compare this measurement-based estimate with the actual (true) proportion of infected among living people,
\begin{equation}
    s(t) = \frac{I(t)}{N'(t)} \, . 
\end{equation}
\begin{equation}
    N'(t) = N- D(t) = S(t)+E(t)+I(t)+R(t)
\end{equation}
is the number of living people in the considered population. Then, the relative error becomes
\begin{equation}
    \epsilon := \frac{|\hat{s}(t)- s(t)|}{|s(t)|} 
    = \left| \frac{[(1-\beta) p I + (1-\beta') q E + \alpha q (S+R)]}{[p I + q (N'-I)]} \, \frac{N'}{I}  - 1\right|
    \, . 
\end{equation}
If the tests measure the Exposed as if they were ``still healthy'', it is reasonable to assume $(1-\beta') = \alpha$.\footnote{because Exposed tested positive would then be considered ``false positives''---they would be valued ``true positive'', once they have transitioned to the ``Infected''} This simplifies the previous formula to
\begin{equation}
    \epsilon = \frac{|\hat{s}- s|}{|s|} 
    = \left| \frac{[(1-\beta) p I + \alpha q (N'-I)]}{[p I + q (N'-I)]} \, \frac{N'}{I}  - 1\right| 
    = \left| \frac{[(1-\beta) p s + \alpha q (1-s)]}{s[p s + q (1-s)]}  - 1\right|
    \, . \label{relerr}
\end{equation}
Let us investigate two limiting cases.
First, if we assume that only people with symptoms are tested, then $q=0$ and
\begin{equation}
    \epsilon = \frac{|\hat{s}- s|}{|s|} 
    = \left| \frac{(1-\beta)}{s} \,  - 1\right|  \, . 
\end{equation}
Accordingly, the smaller the actual proportion $s(t)$ of Infected, the bigger will the relative error be.
Second, if the number $(N'-I) = (S+E+R)$ of people without symptoms typically outweighs the Infected by far, one may assume that $\alpha q (S+E+R) = \alpha q (N'-I) \gg (1-\beta) p I$ and $q (N'-I) \gg p I$. In this case, we get 
\begin{equation}
\epsilon = \frac{|\hat{s} - s|}{|s|} 
\approx \left| \frac{\alpha}{s} - 1\right|\, . 
\end{equation}
\par
In both cases, we can see that the relative error increases with the inverse of the proportion $s$ of Infected. Accordingly, the relative error $\epsilon$ surprisingly increases with the number of healthy people. This relative error might be huge. In fact, it is particularly big in case of ``mild'' epidemics characterized by $s(t) \approx 0$. Again, 
given the finite value of the false positive rate $\alpha$, mass testing of people who feel healthy is not really advised. It might lead to a large overestimation of the actual disease rate. That is, the state of the epidemic may be wrongly assessed.  
\par
\begin{figure}
\parbox[b]{.5\linewidth}{ \includegraphics[width=\linewidth]{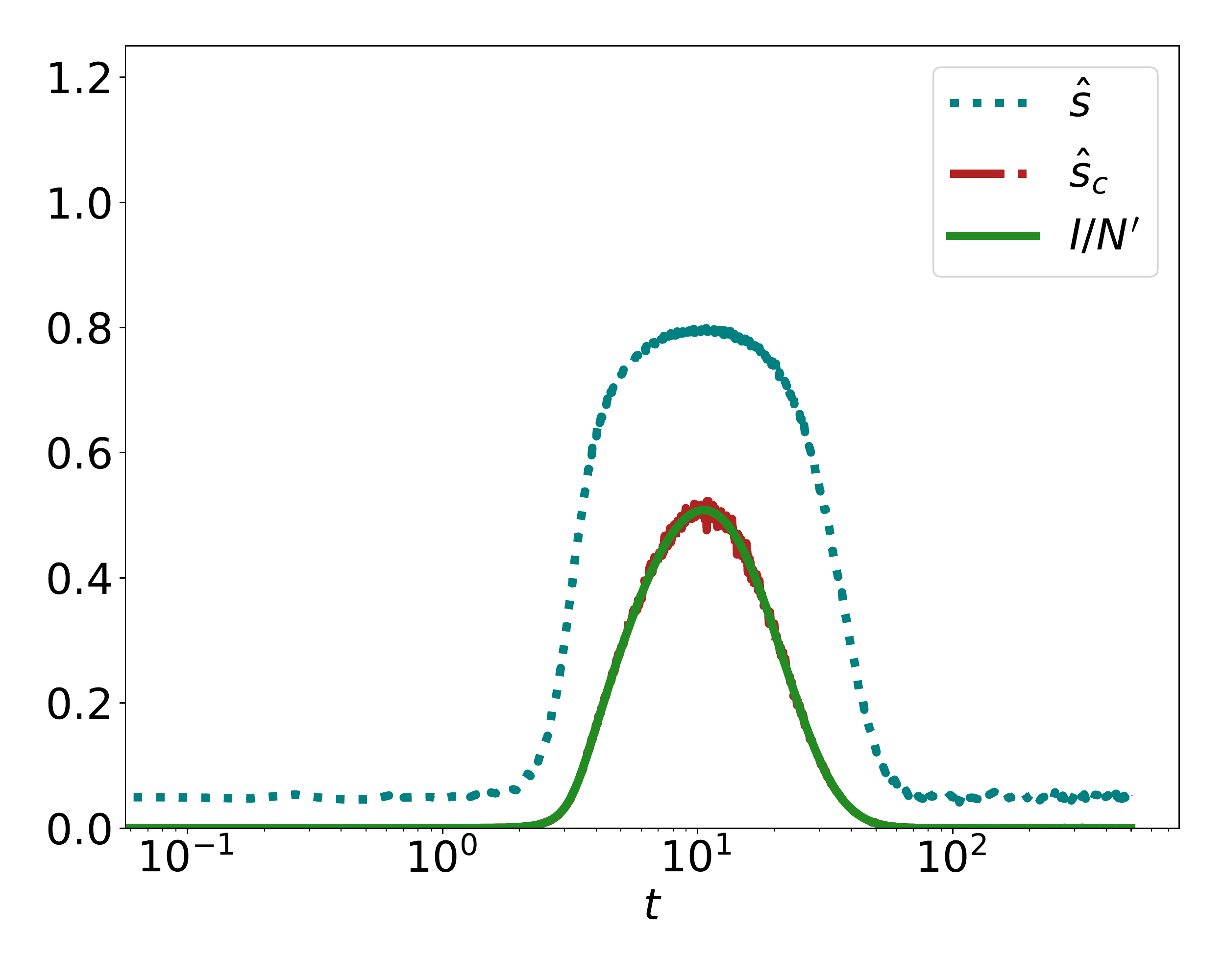}}%
\hspace{.05\linewidth}%
\parbox[b]{.45\linewidth}{\caption{Estimation of the true proportion $I/N'$ of Infected (green solid line) based on the fraction \eqref{success} of positive tests (blue dotted line) and the correction formula \eqref{corrected} (red dashed-dotted line). The corrected estimate tends to be very close to the true value when the parameters $\alpha$, $\beta$, $p$ and $q$ of the test method are known exactly. The curves displayed here are for a SEIRD dynamics with parameters $N=100'000$, $a=1/5$, $b=50$, $c=1/14$, $d=1/14$.\label{Fig2}\vspace*{1cm} } }
\end{figure}
\begin{figure}[ht]
 \centering
    \includegraphics[width=0.4\linewidth]{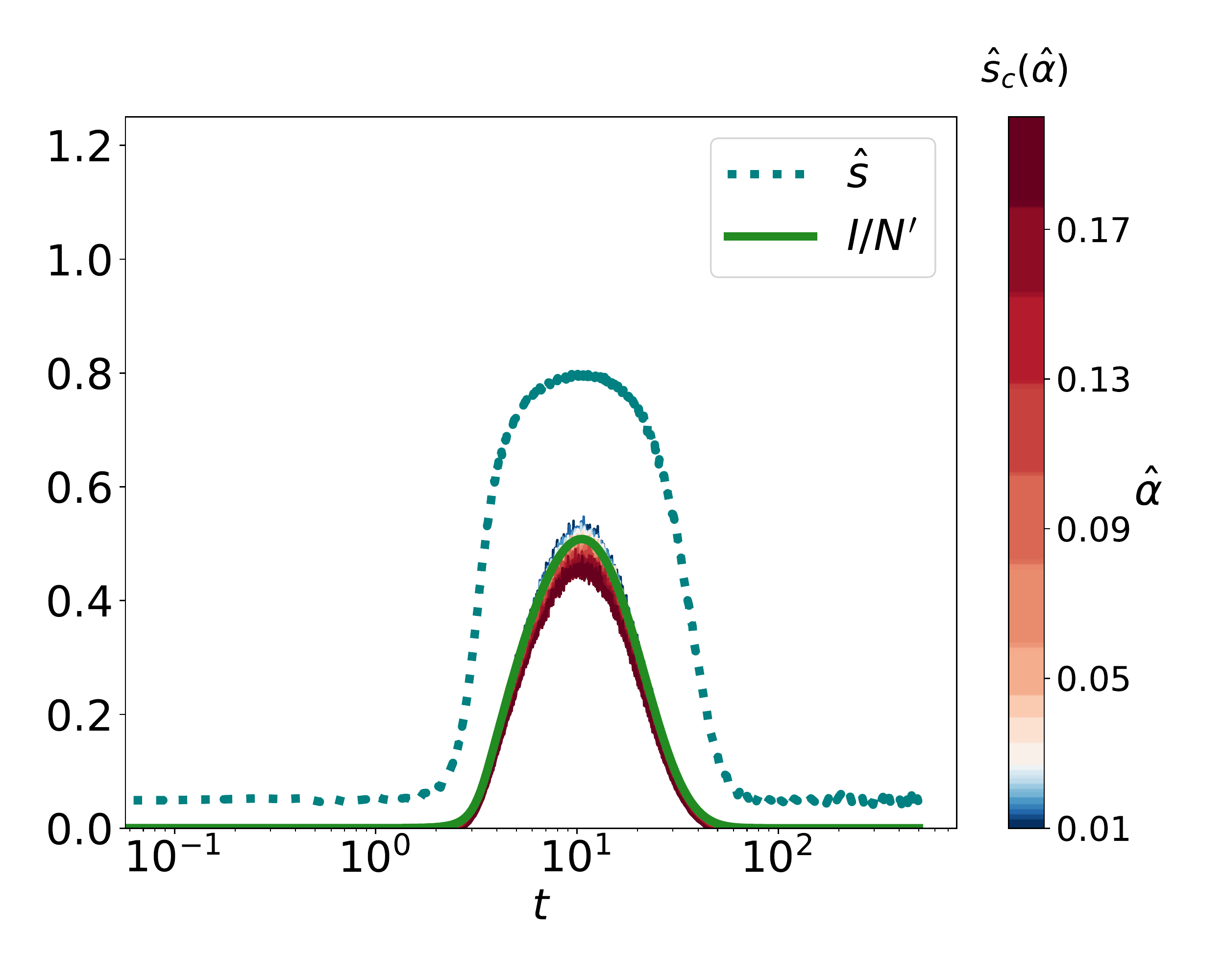}
    \includegraphics[width=0.4\linewidth]{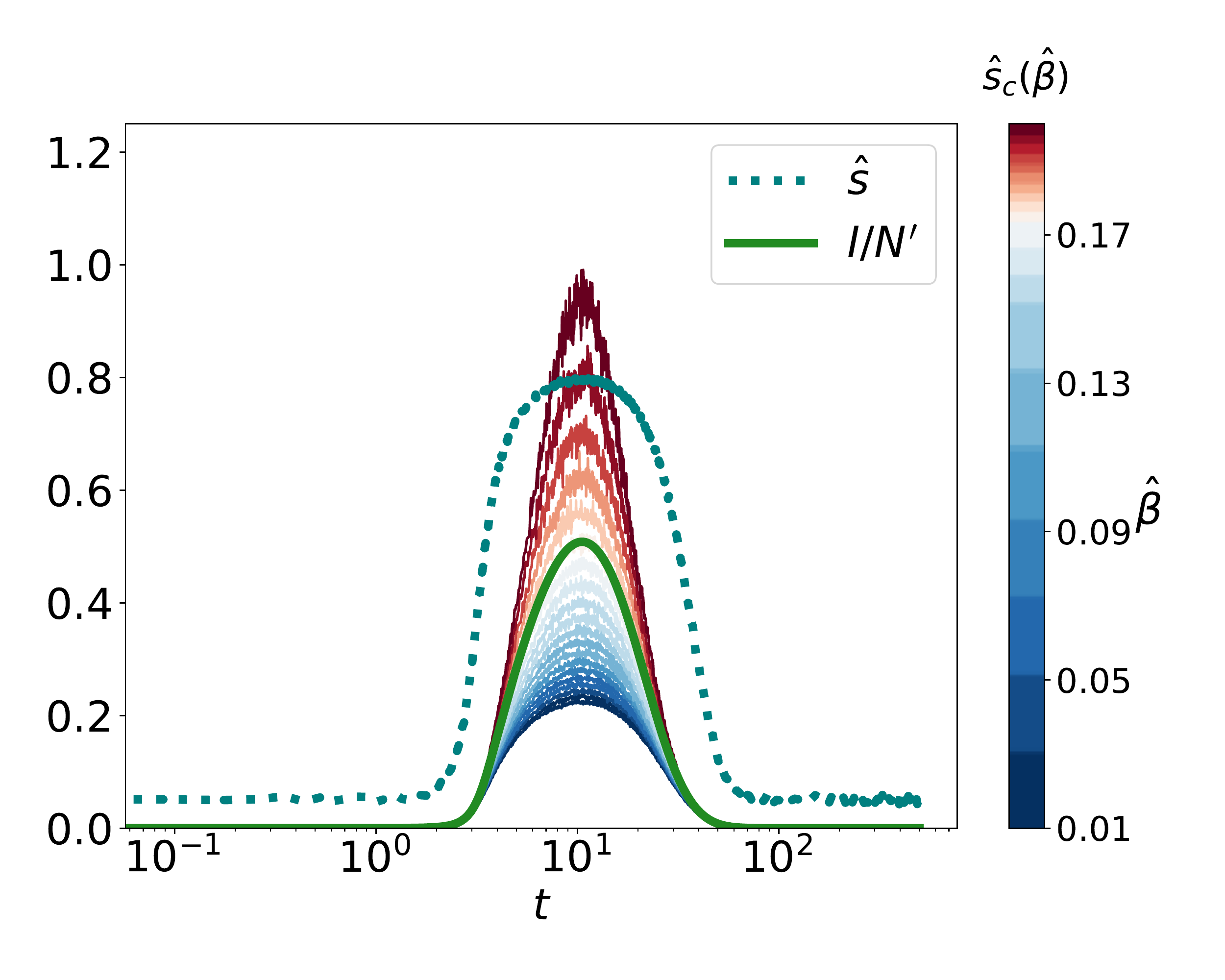} \\[-5mm]
    \includegraphics[width=0.4\linewidth]{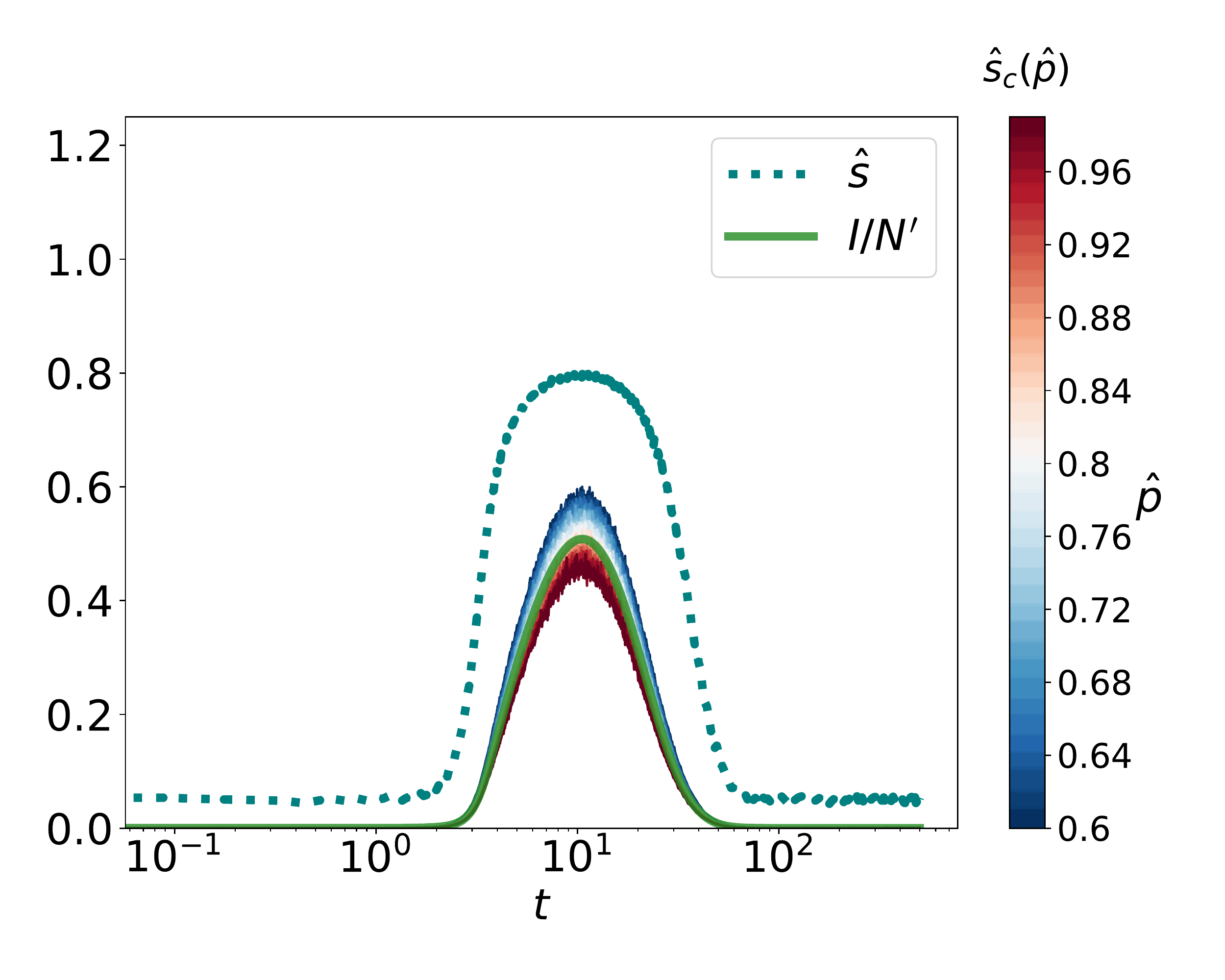}
    \includegraphics[width=0.4\linewidth]{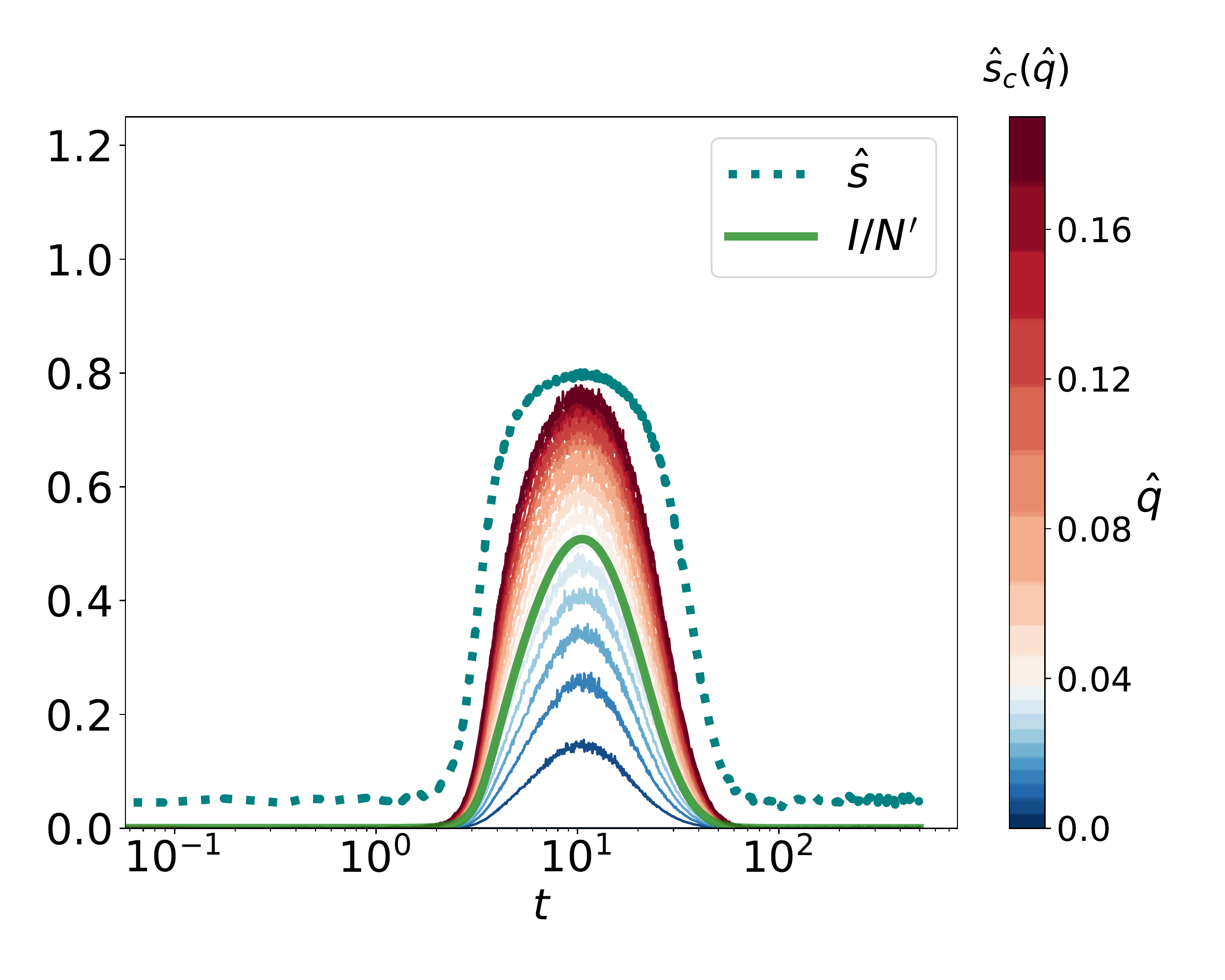}
    \caption{Estimation of the true proportion $I/N'$ of Infected (green solid line) based on the fraction \eqref{success} of positive tests (blue dotted line) and the correction formula \eqref{corrected} (other lines according to the color scales on the right). In these figures we assume that only 3 of the 4 parameters $\alpha=0.05$, $\beta=0.15$, $p=0.8$, and $q=0.06$ are known exactly, while the estimate of the fourth one (represented with a hat on top) is uncertain and, therefore, varied. Specifically, in \eqref{corrected} we have replaced $\alpha$ by $\hat{\alpha}$ in the top left figure, $\beta$ by $\hat{\beta}$ in the top right figure, $p$ by $\hat{p}$ in the bottom left figure and $q$ by $\hat{q}$ in the bottom right figure. All figures show results for a SEIRD dynamics with $N=100'000$, $a=1/5$, $b=50$, $c=1/14$, $d=1/14$.}
    \label{fig_s_correction} \label{Fig3}
\end{figure}

However, there is a correction formula for the effect of false positives and negatives, and for biased samples with $p \ne q$. Similar to the error and bias correction introduced in Ref.~\cite{bottcher2021TestStatistics}, to estimate the fraction of unhealthy people correctly, we need to find a function $\hat{s}_c$, which transforms the estimate $\hat{s}$ into $s$. 
Considering that the estimate $\hat{s}$ of $s$ is given by \eqref{success}, and
$s = I/N' = I / (S+E+R)$, where $N' = (S+E+I+R)$, we find
\begin{equation}
    \hat{s}(s)  
    = \frac{(1-\beta) p s  + \alpha q (1-s)}{(p-q) s + q}  
    \label{hats}
\end{equation}
under the previously made assumption $(1- \beta') = \alpha$.
From this, we can derive the corrected value $\hat{s}_c$ of $\hat{s}$ via the inverse function of $\hat{s}(s)$:
\begin{equation}
     \fbox{ $ \displaystyle \hat{s}_c (\hat{s}) := s(\hat{s}) = \frac{(\hat{s}-\alpha)q}{(1-\beta)p - \alpha q - (p-q)\hat{s}} $}
\label{corrected}
\end{equation}
This formula should only be used in the range $0 < \hat{s}-\alpha < 1 - \alpha -\beta$.
For non-biased samples ($p=q$), the formula simplifies to 
\begin{equation}
    \hat{s}_c (\hat{s}) = \frac{\hat{s} - \alpha}{1 - \alpha - \beta}
\end{equation}
As \figref{Fig2} shows, this correction can be very effective in fitting the true value $I/N'$, if the parameters $\alpha$, $\beta$, $p$ and $q$ are known. 
However, we would like to point out that the above analysis and correction formula are based on a mean-value approximation. Let us, therefore, investigate the likely role of stochastic measurement errors. A common approach for this is the use of a binomial distribution. It describes the probability to have $n_p$ positive tests among $N_T$ tests, if the ``success rate'' (of having a positive test result) is $\hat{s}$. Accordingly, the binomial distribution reads
\begin{equation}\label{eq:nc_noerror}
    P(n_p|N_T)= \binom{N_T}{n_p}\hat{s}^{n_p}(1-\hat{s})^{N_T-n_p} \, , 
\end{equation}
where $N_T = pI + q(S+E+R)$ represents again the total number of tests made.
For the binomial distribution, the expected number of confirmed cases is
\begin{equation}
    \langle n_p \rangle = N_T \, \hat{s} = N_p \, , 
\end{equation}
and the variance is
\begin{equation}
    \mbox{Var}(n_p) = N_T \, \hat{s} (1-\hat{s}) \, . 
\end{equation}
As a consequence, the relative standard deviation is
\begin{equation}
    \frac{\sqrt{\mbox{Var}(n_p)}}{\langle n_p \rangle}
    = \frac{\sqrt{N_T \, \hat{s} (1-\hat{s})}}{N_T \, \hat{s}}
    = \frac{1}{\sqrt{N_T}} \sqrt{\frac{1-\hat{s}}{\hat{s}} } \, . 
\end{equation}
Here, more tests are better, as expected. While the relative standard deviation may be significant if $N_T$ is small (in the beginning of an epidemics) or if $\hat{s}\approx 0$, for a large enough number of tests, the relative variance will be rather negligible, say, of the order of 1\%. Therefore, the above mean value equations and correction formula \eqref{corrected} appear to be applicable in many cases, when the uncertainty in $\alpha$, $\beta$, $p$ and $q$ is low.
In reality, however, the parameter values showing up in formula \eqref{corrected} are estimates $\hat{\alpha}$, $\hat{\beta}$, $\hat{p}$ and $\hat{q}$, which may deviate from the true parameter values $\alpha$, $\beta$, $p$, and $q$. Inserting these in formula \eqref{corrected} instead, one may find considerable under- or over-estimations of the true proportion of Infected in a population (see \figref{fig_s_correction}). 
\par
A further complication occurs, if the assumption $(1-\beta') = \alpha$ does not hold (e.g.\ when the testing procedure recognizes Exposed as ``already infected'' people, even though they do not have symptoms yet). Then, it is generally not possible to derive a correction formula such as \eqref{corrected}. To derive such a formula, one would need to separately measure Infected and Exposed, requiring two different, specific tests, but this is often not realistic to assume. One would have to use additional assumptions or Bayesian inference (see below). 
\par
Last but not least, network effects may also produce significant deviations from the mean-value approximations above (which implicitly assumed homogeneous mixing and a low variance of measurement errors). To assess their size, in the following section we will use a more sophisticated simulation approach for the SEIR/SEIRD model that considers stochastic as well as network effects.

\section{STOCHASTIC SIMULATION OF EPIDEMIC SPREADING IN SOCIAL NETWORKS}\label{sec:model}

Let us start by discussing the full stochastic SIR model on a contact network $G=(V,E)$, defined by a set of nodes $V$ (corresponding to the individuals) and a set of edges $E$ (representing their contacts). The SIR process is determined by parameters  $(b, c)$, where a susceptible individual becomes exposed with rate $b$ by having a contact with an infected neighbour, and an infected node recovers or dies with a rate $c+d$.  
These processes have exponential inter-event time distributions $\psi(\tau)=b e^{-b \tau}$ (spreading) and $\phi(\tau)=(c+d) e^{-(c+d) \tau}$ (recovery or death).
\par
For a given contact network $G$ and the stochastic SIR epidemic spreading model, we create weighted networks $\left\lbrace G_k \right\rbrace$ and simulate realizations of the stochastic spreading dynamics on them. Each weighted network $G_k$ is linked to a possible outcome of the epidemic spreading process, starting from a randomly selected ``source'' node. In order to extract samples of epidemic trajectories of the full stochastic SIR model on a contact network, we use a SP-KMC shortest-path Kinetic Monte Carlo method \cite{PhysRevResearch.2.033121,tolic2018simulating}. A \textit{time-respecting weighted network} instance $G_k$ is created by taking the input network $G$ and assigning weights to the edges of the network according to 
\begin{equation}
\label{eq:Poissonweights}
\rho_{i,j} =
\begin{cases} 
      -\ln(x)/b & \mbox{ if }-\ln(x)/b \leq -\ln(y)/(c+d) \\
      \infty & \mbox{if } -\ln(x)/b > -\ln(y)/(c+d) \\
   \end{cases}
\end{equation}
Here, $x,y$ are uniform random numbers $\in [0,1]$. 

\begin{figure}[ht]
    \centering
    \includegraphics[width=0.45\linewidth]{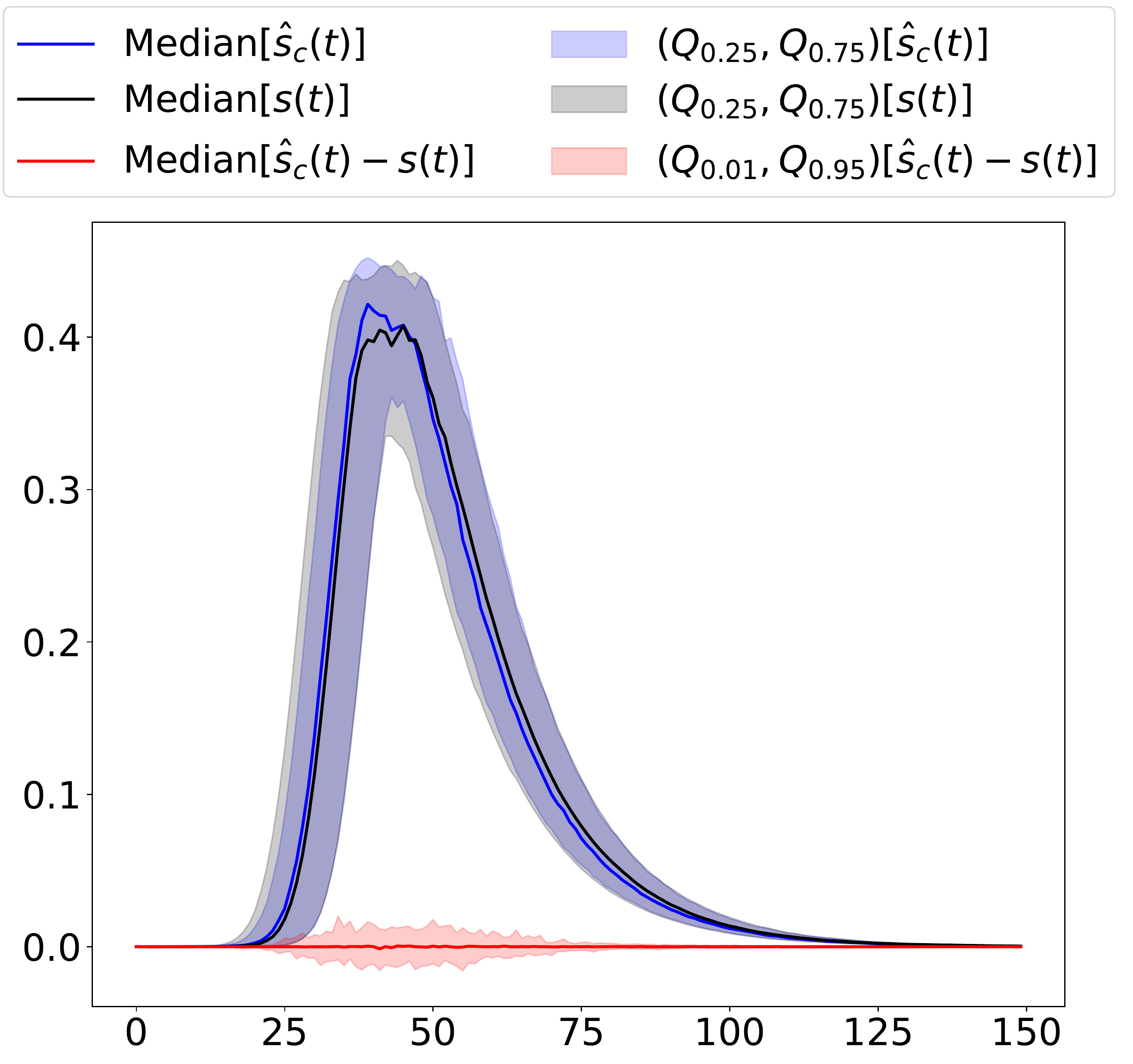}
    \includegraphics[width=0.45\linewidth]{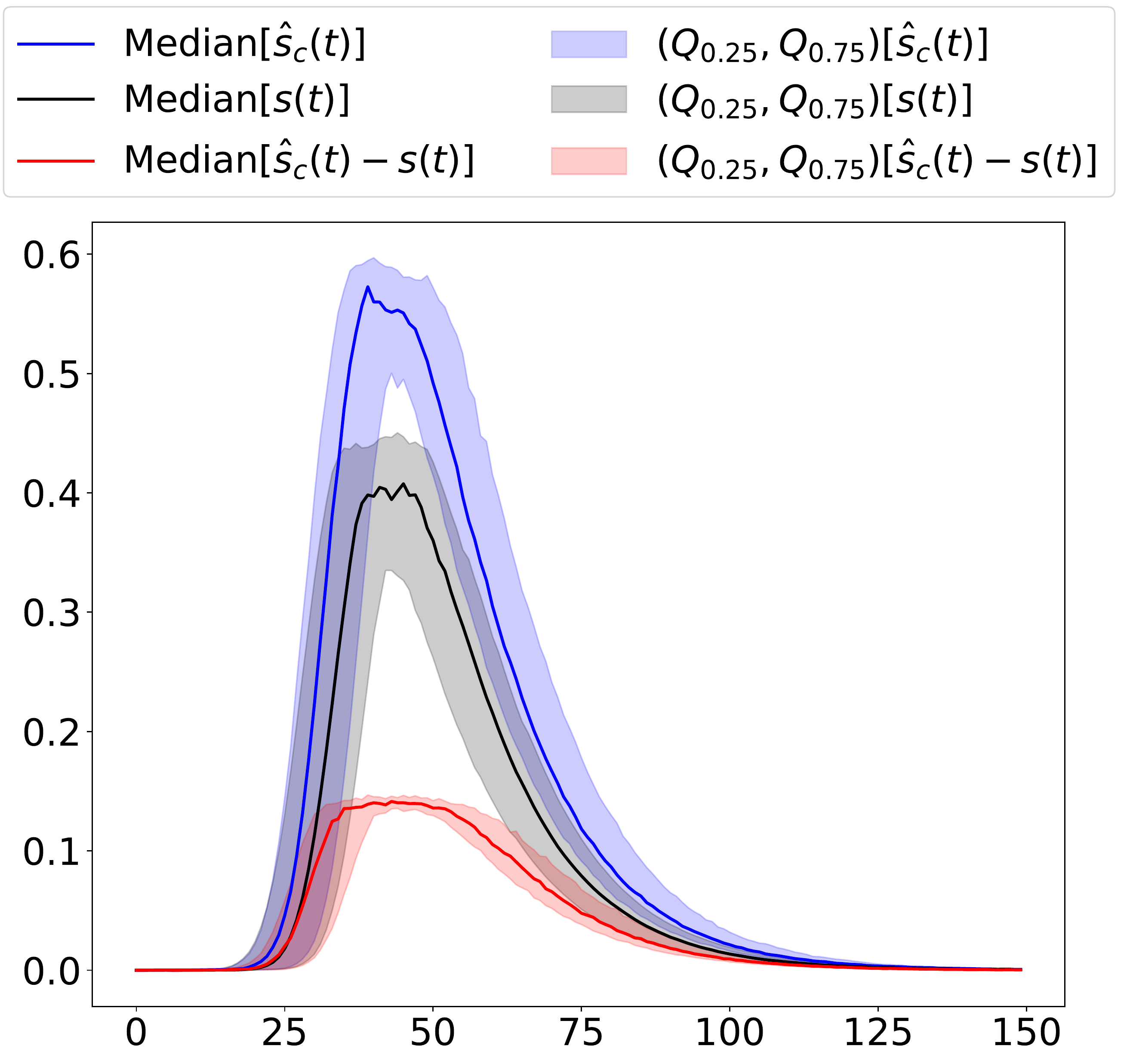}
    \caption{Ensemble of stochastic trajectories (SP-KMC sampling) on an ensemble of $10^3$  Barab\'asi-Albert networks with $N=100'000$, $m=3$, and $\gamma=1$. 
    Left: Scenario assuming that the parameters $\alpha=0.05$, $\beta=0.15$, $p=0.8$, $q=0.06$ of the test method are exactly known. 
    Right: Reconstruction for somewhat incorrect estimates $\hat{p}=0.75$ and $\hat{q}=0.1$, while the other parameters are assumed to be the same. The assumed parameters of the SEIRD dynamics are $a=1/5$, $b=3/14$, $c=1/28$, $d=1/28$. $Q_{0.01}$, $Q_{0.25}$, $Q_{0.75}$  and $Q_{0.99}$ represent 1, 25, 75, and 99 percent quantiles. The median values and error quantile bands are based on $10^3$ simulations.}\label{fig:BA_mean_field_ensemble}\label{Fig4}
\end{figure}

In order to work with the stochastic epidemic SEIR model, we extend the SP-KMC mapping by considering transitions from the Exposed state to Infected state with rate $a$. This process has the exponential inter-event incubation time distribution $\chi(\tau)= a e^{-a \tau}$. 
In particular, a
\textit{time-respecting weighted network} instance $G_k$ is created by taking the input network $G$ and assigning 
weights to the edges of the network instance as
\begin{equation}
\label{eq:Poissonweights}
\rho_{i,j} =
\begin{cases} 
      -\ln(x)/b  -\ln(z)/a & \mbox{if } -\ln(x)/b \leq -\ln(y)/(c+d) \\
      \infty & \mbox{if } -\ln(x)/b > -\ln(y)/(c+d) \\
   \end{cases}
\end{equation}
Here, $x,y,z$ are uniform random numbers $\in [0,1]$. 
We define the distance as shortest path on a weighted networks:
\begin{equation}
\mathit{d}_{G_k} (v_i,v_j) = \min_{\chi_{ij}} \sum_{(k,l)\in\chi_{ij}} \rho_{k,l} \, , 
\end{equation}
where $\chi_{ij}$ is the set of all possible paths from node $v_i$ to node $v_j$ on network $G_k$ and $\rho_{k,l}$ denotes the weights defined in~\eqref{eq:Poissonweights}. Now, epidemic properties can be extracted~\cite{PhysRevResearch.2.033121,tolic2018simulating} from the node and edge weighted networks $\left\lbrace G_k \right\rbrace$. 
The run-time complexity of extracting a single epidemic trajectory is dominated by Dijkstra's shortest paths algorithm from a specific source node to others. It is of the order $O(L+M\log M)$, where $L$ denotes the number of edges in the network and $M$ the number of nodes. 

 \figref{fig:BA_mean_field_ensemble} shows on the left that the mean-field correction \eqref{corrected} works well if all measurement parameters $\alpha$, $\beta$, $p$, and $q$ are known. However, for somewhat incorrect parameter estimates, we observe considerable deviations from the mean-field correction. This establishes the need for Bayesian inference.

\begin{figure}
    \centering
    \includegraphics[width=0.38\linewidth]{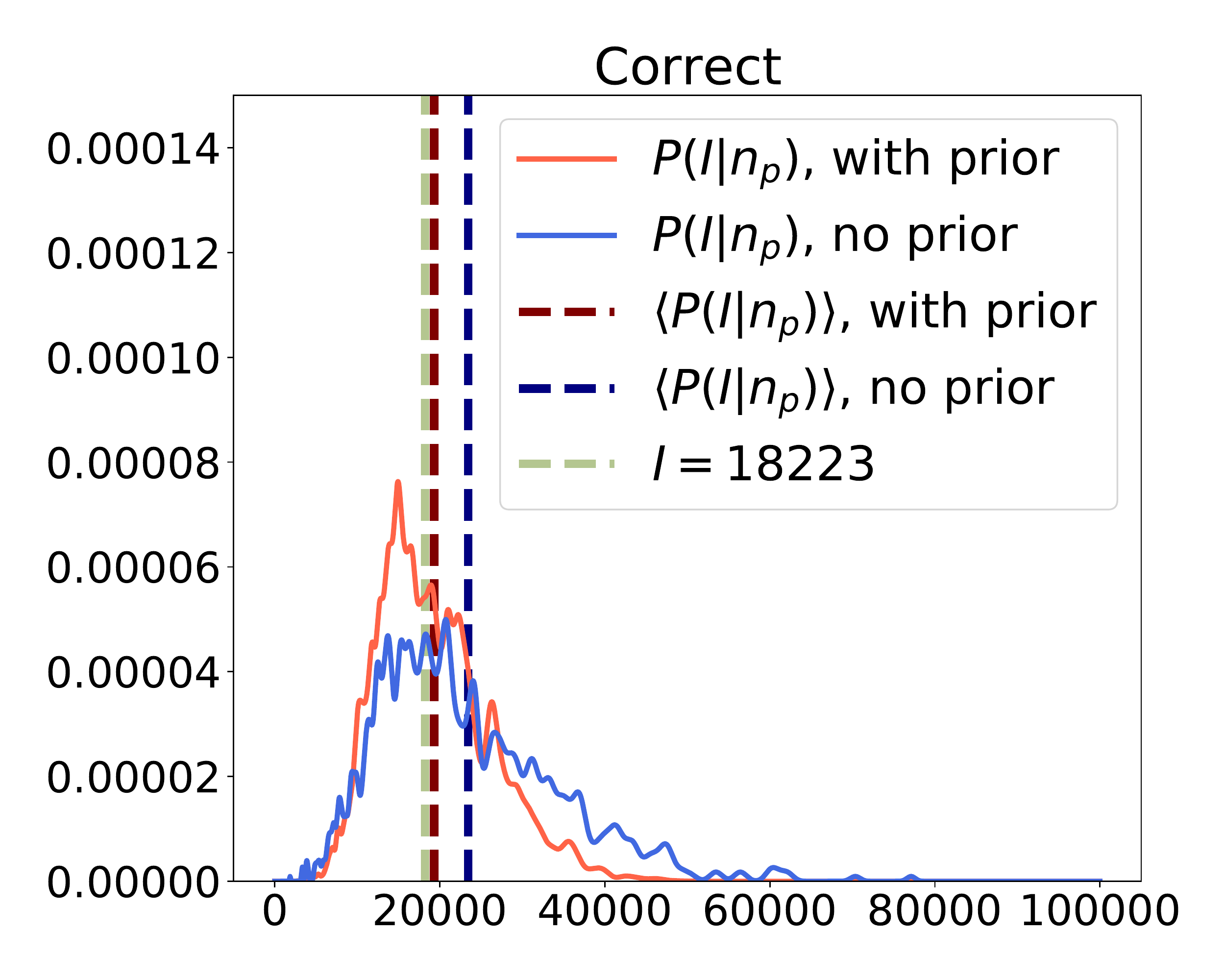}
    \includegraphics[width=0.38\linewidth]{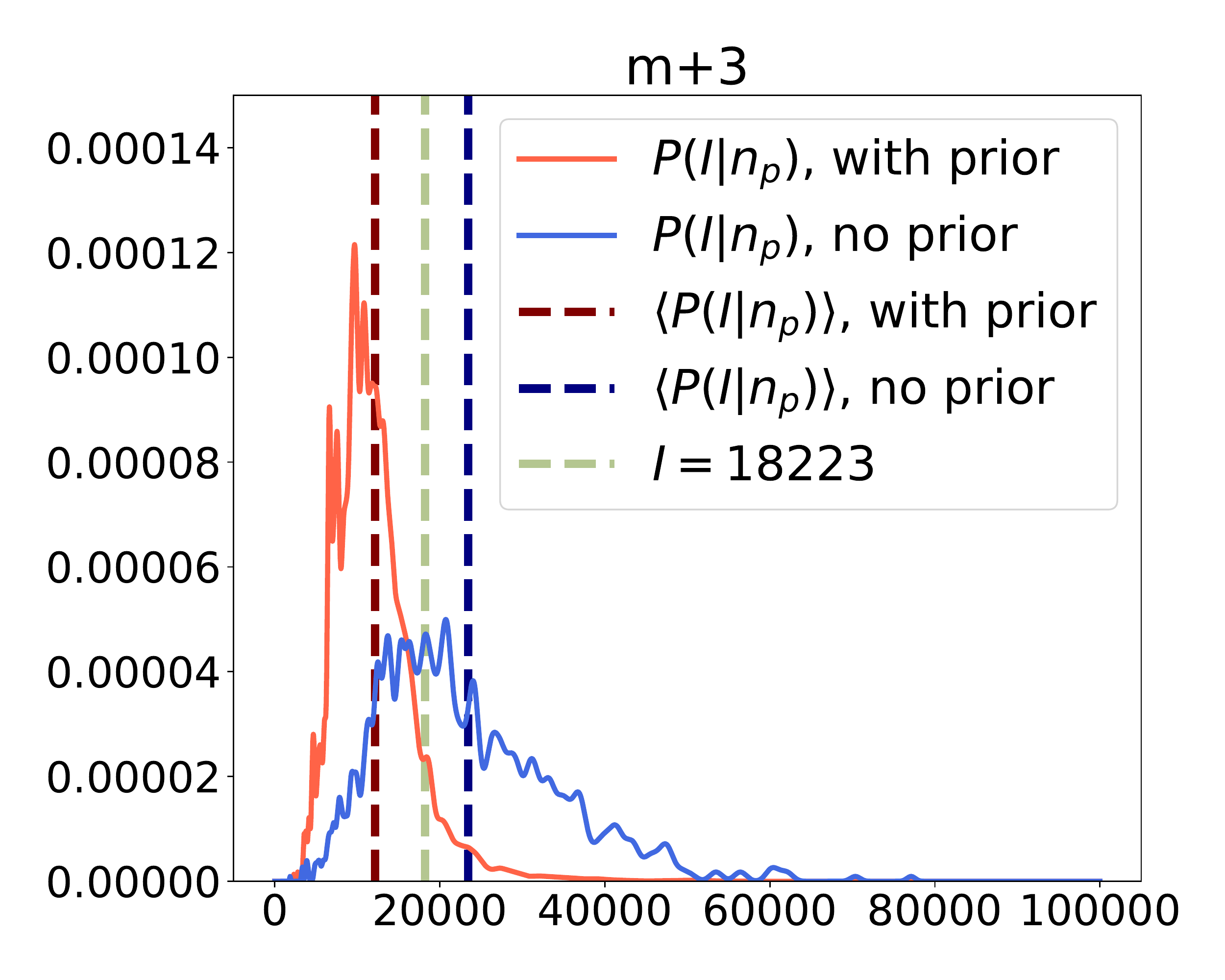}
    \\
    \includegraphics[width=0.38\linewidth]{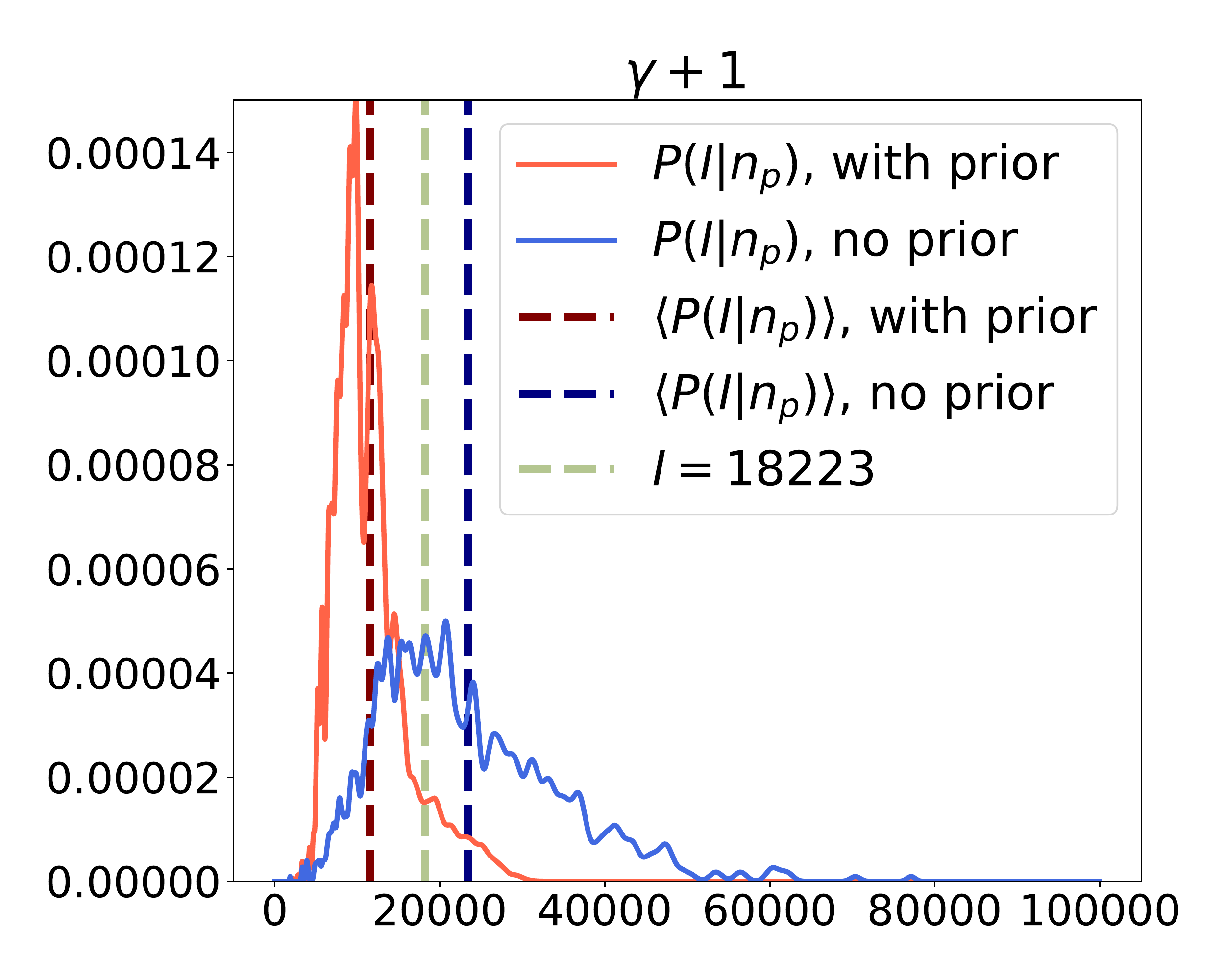}
    \includegraphics[width=0.38\linewidth]{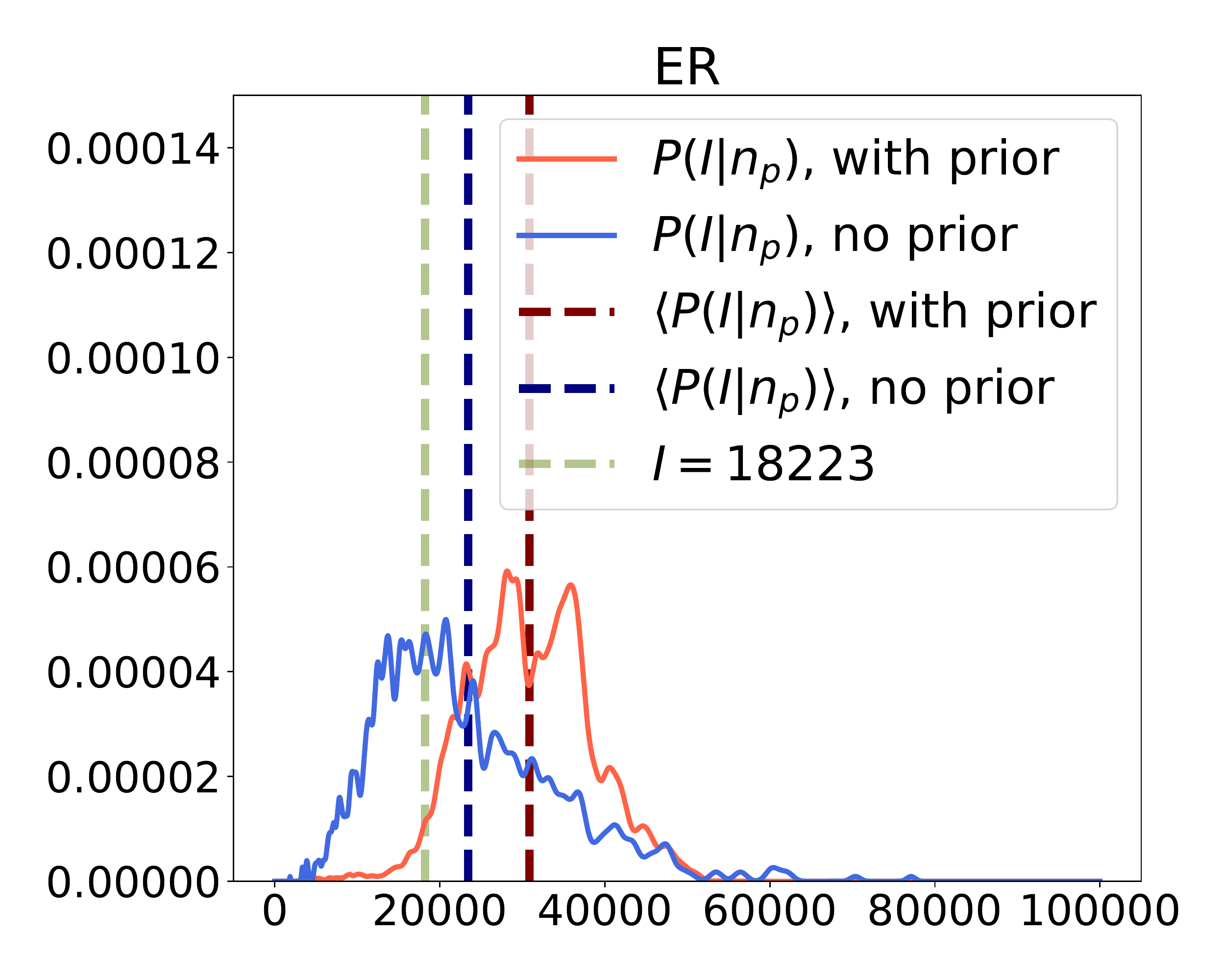}
    
    \caption{Posterior distribution $P(I|n_p)$ when a network topology prior is used (in red) and when not (in blue). In scenario ``Correct'' we used a correct network ensemble contact prior for Barab\'asi-Albert networks with $N=100'000$, $m=3$, and $\gamma=1$, whereas in ``m+3'' we used a network ensemble contact prior for Barab\'asi-Albert networks with $m=6$ and $\gamma=1$. In ``$\gamma+1$'' we used a prior for Barab\'asi-Albert networks with $m=3$ and $\gamma=2$, and in ``ER'' we used a prior for Erd\"os-R\`enyi networks with $\lambda=6$, which is far from the true degree distribution. In all figures, the green line shows the ``ground truth'', against which the biased testing is performed. We observe that knowledge about the degree distribution of contacts helps to estimate the true ground truth. To estimate network priors within $t \in [60,65]$, we have used 100 networks from the ensemble.\label{fig_bayes_networkpriors}\label{Fig5}}
\end{figure}

If the bias parameters $p$ and $q$ are uncertain, one can use the prior distribution $\hat{p},\hat{q} \sim P(p,q)$
and integrate \eqref{eq:marginalization} over all possible values to obtain posterior distribution $P(I|N_p=n)$.
Then, the corrected estimate of the number of infected is given by the expected posterior
\begin{equation}
    \hat{I} =  \langle P(I=m|N_p=n) \rangle \, , 
\end{equation}
where $\langle . \rangle$ is the expected value of posterior $P(I=m|N_p=n)$, which is estimated using the Bayesian formula \eqref{eq:bayes}. In the absence of knowledge about the epidemic dynamics or network, one may use the uninformative prior. However, if there exists knowledge about the degree distribution $P(k)$ of network contacts, one can estimate $P(I_t=m)$ through the use of SP-KMC sampling on the network ensemble with a given degree distribution $P(k)$.

In Fig. \ref{fig_bayes_networkpriors} we show that, using stochastic trajectories generated with networks from a correct ensemble of networks, one can improve the estimation of $\hat{I}=\hat{s}N'$. However, when using a prior with wrong assumptions about the degree distribution of contacts, we see no improvement in the posterior estimate of the ground truth, compared to the uninformative prior. Overall, we find that both the degree distribution and the network density have effects on the quality of the Bayesian inference. Therefore, we would like to emphasise the importance of choosing correct priors for a reliable estimation of the ground truth.

\section{Summary, Conclusion, Discussion, and Outlook}
In this paper, we have addressed some challenges regarding epidemic monitoring and modelling, particularly regarding (i) the modeling itself and (ii) the applicability of models to the real world. 
Regarding the first point, it matters to choose the right epidemiological model, but also to consider limitations of the common mean value approach. Stochastic and network effects---in combination with the underlying non-linear spreading dynamics---are relevant as well. Regarding the second point, one needs to consider errors of tests and estimation procedures underlying epidemic monitoring. This concerns false positive and false negative rates, but also biases in the measurement sample ($p\ne q$). Our goal was to analyse a good approximation of reality (where we have represented the latter by an assumed ground truth, specifically a SEIRD dynamics). Overall, we showed that data-driven approaches that rely only on reported infected cases can be misleading, if reliable measurement corrections are not available or used. 
\par
Monitoring real-world epidemics is difficult, because one does not know the ground truth, and the goal of all types of modelling is to infer it. Although a simple mean-field correction works well when the measurement parameters are known (see Fig. \ref{Fig2}), substantial deviations may occur when there is some uncertain about them (see Fig. \ref{Fig3}), which is the usual case. Then, one might resort to a Bayesian correction for improvements~\cite{bottcher2021TestStatistics,campbell2020bayesian,bentley2021error,catala2021robust,wu2020substantial}, but this will also not remove all issues of measurement errors and uncertainties completely. 
\par
For all the above reasons, forecasting the epidemic dynamics is a problem that has fundamental limitations even in the age of Big Data and Artificial Intelligence. Given the intrinsic limitations of tests, these problems are expected to stay around. Measurement problems using dynamical models~\cite{yang2014comparison, engbert2021sequential} may be further amplified if the test method or the social behavior of people are dynamically changed in response to the epidemic dynamics. Hence, data protection regulations are not the main problem for monitoring, forecasting and controlling epidemics. The limitations of a data-driven approach are a lot more fundamental in nature.
\vskip6pt

\enlargethispage{20pt}

{\it Data Accessibility:} This paper is not presenting any empirical data.\\[3mm]
{\it Author Contributions:} All authors developed the theoretical concepts and experiments together. NAF and VV created the computer code used and performed the simulations. All authors wrote the manuscript. DH initiated and coordinated the study. All authors read and approved the final manuscript.\\[3mm]
{\it Competing Interests:} The authors declare no competing interests.\\[3mm]
{\it Funding:} The authors acknowledge support by the SOBIGDATA++ project funded by the European Union’s Horizon 2020 research and innovation programme under grant agreement No. 871042.\\[3mm]
{\it Acknowledgments:} We would like to thank Lucas Böttcher for inspiring discussions and scientific exchange regarding models of statistical measurement~\cite{bottcher2021TestStatistics} in the initial phase of the project.

\appendix

\section{Bayesian Inference for $\hat{I} = \hat{s}N'$}

In the main text of this paper, we have assumed that the testing rates $p$ and $q$ are fractions determined by the health system. However, if they are considered to reflect the probability of infected and healthy individuals to get tested, they may be considered to be random variables that vary from one individual to another. For example, we may assume $p$ and $q$ to be Beta-distributed according to probability distributions $P(p)=\textrm{Beta}(A_I,B_I)$, $P(q)=\textrm{Beta}(A_H,B_H)$. Considering that the number of Infected and the number of positively tested people are also random variables in reality, we may use Bayes' rule, where our evidence is $N_p$ and our priors relate to knowledge about $P(I)$. Bayes' rule determines the probability that the hypothesis is true (here: the number of Infected $I$ is $m$) given a certain observation (here: the measured number $n$ of positive tests): 
\begin{equation}
\label{eq:bayes}
    P(I=m|N_p=n) = \frac{P(N_p=n|I=m)P(I=m)}{\sum_l P(N_p=n|I=l)P(I=l)}.
\end{equation}
Here, the left-hand side of the equation is the ``posterior'' probability that we aim to infer. The best estimate for $I$ is given by the maximum likelihood estimate $\hat{I}$, i.e.\ by the value $I=m$ for which the for which $P(I=m|N_p=n)$ is maximum: 
\begin{equation}
    \hat{I} = \underset{m\in [0,N']}\argmaxA P(I=m|N_p=n) \, . 
\end{equation}

Here, 
\begin{equation}
    \label{eq_bayes}
    P(I=m|N_p=n)~\sim  \frac{1}{M}\sum_{i=1}^M \left( \frac{1}{\sqrt{2\pi N_T \hat{s}_i(1-\hat{s}_i)}} \exp \left( \frac{-(n-N_T\hat{s}_i)^2}{2 N_T \hat{s}_i(1-\hat{s}_i)} \right) \right)
\end{equation}
with $\hat{s}_i = \hat{s}(p_i,q_i)$ for each sample $i$ from the prior distribution. Eq.~\eqref{eq_bayes} is a practical extraction of the posterior probability distribution using Monte Carlo sampling, making the assumption that the Binomial distribution discussed before can be approximated by a normal (Gaussian) distribution, which is often the case. 
Overall, we find the Bayesian estimate
\begin{equation}
     \boxed{\hat{I} = \langle P(I=m|N_p=n) \rangle=\sum_{m=0}^N
     \frac{m}{M}\sum_{i=1}^M \left( \frac{1}{\sqrt{2\pi N_T \hat{s}_i(1-\hat{s}_i)}} \exp \left( \frac{-(n-N_T\hat{s}_i)^2}{2 N_T \hat{s}_i(1-\hat{s}_i)} \right) \right) } 
\label{eq_bayes_2}
\end{equation}
where $ \langle . \rangle$ denotes the expectation over the values of $P(I=m|N_p=n)$.
\par
Finally, let us present the full derivation of \eqref{eq_bayes}. As mentioned above,
Bayes' theorem \eqref{eq:bayes} allows one to calculate the probability that the number $I$ of Infected is $m$, given a measured number $n$ of positive tests.
For example, if we assume an uninformative prior with $P(I=m)=c$  $\forall \, m$, then, up to a normalisation factor, we can consider $P(I=m|N_p=n)\propto P(N_p=n|I=m)$. 

Using Bayes' equation
\begin{equation}
    P(m|n,p,q) =\frac{ P(m,n,p,q)}{P(n,p,q)}=\frac{P(n,p,q|m)P(m)}{\sum_{m'} P(n,p,q|m')P(m')}
\end{equation}
and considering 
\begin{equation}
\label{eq:marginalization}
 P(N_p = n|I= m)  =  \int_0^1 \int_0^1 P(N_p=n,p=p_j,q=q_j|I=m) dp_j dq_j \, ,  
\end{equation}
\begin{equation}
P(N_p=n,p,q|I=m)P(I=m) = P(N_p=n|I=m,p,q)P(I=m)P(p,q) \, , 
\end{equation}
gives the multivariate formula
\begin{equation}
P(I=m|N_p=n) = \frac{ \int_0^1 \int_0^1 P(N_p=n|I=m,p=p_i,q=q_i)P(I=m)P(p_i,q_i) dp_i dq_i }{\sum_{l} \int_0^1 \int_0^1 P(N_p=n|I=l,p=p_i,q=q_i)P(I=l)P(p_i,q_i) dp_i dq_i } \, . 
\end{equation}
For an uninformative prior for $P(m)$ we have 
\begin{equation} 
P(I=m|N_p=n)\propto  \int_0^1 \int_0^1 P(N_p=n|I=m,p=p_i,q=q_i)P(p_i,q_i) dp_i dq_i \, . 
\end{equation}
In case of Beta priors for $p$ and $q$, we have
\begin{equation}
    P(p,q)=\textrm{Beta}(A_I, B_I)(p) \, \textrm{Beta}(A_H, B_H)(q) \, ,
\end{equation}
where $\textrm{Beta}(A, B)(x)= \frac{\Gamma(A+B)}{\Gamma(A)\Gamma(B)}x^{A-1}(1-x)^{B-1}$.
We will assume the Gaussian approxi\-mat\-ion:
\begin{equation}
P(N_p=n|I=m,p,q)\approx \frac{1}{\sqrt{2\pi N_T \hat{s}(1-\hat{s})}} \exp \left( \frac{-(n-N_T \hat{s})^2}{2 N_T \hat{s}(1-\hat{s})} \right)  \, , 
\end{equation}
where 
\begin{equation}
    \hat{s} = \hat{s}(I=m,p,q) = \frac{(1-\beta)\, p \, m/N' + \alpha \, q \, (1 - m/N')}{(p-q)\, m/N' + q}  
\end{equation}
in accordance with \eqref{hats}. 
Finally, $P(I=m|N_p=n) \sim $
\begin{equation}
  \int_0^1 \int_0^1 \frac{1}{\sqrt{2\pi N_T \hat{s}(1-\hat{s})}} \exp \left( \frac{-(n-N_T \hat{s})^2}{2 N_T \hat{s}(1-\hat{s})} \right)\textrm{Beta}(A_I, B_I)(p_i) \textrm{Beta}(A_H, B_H)(q_i)dp_i dq_i \, . 
\end{equation}
The integral can be efficiently estimated using Monte Carlo sampling. In that case the integrals turn into a sum:
\begin{equation}
 P(I=m|N_p=n) \sim 
  \frac{1}{M}\sum_{i=1}^M \left( \frac{1}{\sqrt{2\pi N_T \hat{s}_i(1-\hat{s}_i)}} \exp \left( \frac{-(x-N_T \hat{s}_i)^2}{2 N_T \hat{s}_i(1-\hat{s}_i)} \right) \right) \, ,
\end{equation}
where, in order to estimate $\hat{s}_i = \hat{s}(p_i,q_i)$, one selects a sample of $M$ Beta-distributed values $p_i \sim \textrm{Beta}(A_I, B_I)$ and $q_i \sim \textrm{Beta}(A_H, B_H)$. 

In the case of \textit{network priors} $P(I_t=m)$, we consider samples of epidemic trajectories from $N_{\textrm{G}}$ randomly generated networks with given network characteristics. For each of the networks, we simulate $N_e$ epidemic trajectories, each with a randomly chosen initial infected node. The prior distribution $P(I_t=m)$ is then estimated as the probability density function of trajectories in the time window $[t-\delta,t]$. 

\bibliographystyle{unsrtnat}

%
\end{document}